\documentstyle[graphicx,lcy,amsfonts,amssymb,12pt]{article}

\begin{document}

\title{Molecular dynamic investigations of liquid nickels
structural properties}

\author{A.~Kimmel*, V.~Ladyanov*, I.~Gusenkov**}
\maketitle

\begin{center}
{\small \it *~Physical Technical Institute of Ural branch of
Russian Academy of Sciences, 132,~Kirov
Str.,~Izhevsk,~Russia,~426000\\
**~Udmurt State University, 1,~Universitetskaya Str.,~Izhevsk,
Russia,~426034\\}
\end{center}


\sloppy

\begin{abstract}
The effective pair potentials of liquid nickel have been derived
by means of the inverse self-consistent method from the
experimental diffraction data. The potentials provide highly
accurate coincidence of experimental and simulated structural
properties. The statistical analysis of Voronoi polyhedral and
Delaunau simplices has shown that in the vicinity of the melting
point in the structure of liquid nickel there are relatively
stable spatial polytetrahedral formations with local 5-folder
symmetry. With temperature rise such formations evolve to isolate
and compact ones with preserving of the local five-folder
symmetry.
\end{abstract}

\begin{section}{Introduction}

In the last few years, extensive studies have been performed on a
variety of ionic, covalent, and metallic liquids in order to
understand their structural and dynamic behavior from a
microscopic perspective. Transition metals, such as nickel, are
very interesting in this record. The structure of liquid nickel
has been studied in many works by means of direct diffraction
methods and investigations of structural sensitive properties
\cite{Baum, Vatolin, Schenk}. Significant modifications of
structure parameters were obtained above the nickel melting point.
Authors of the work \cite{Vatolin} have obtained the fuzzy maximum
located close to the line (200) of the hypothetical BCC structure.
However, by means of the levitation techniques in conjunction with
the neutron scattering experiment the authors of the work
\cite{Schenk} unambiguously proved the Franks hypothesis
\cite{Frank}, which states that the icosahedral short range order
is already revealed above the melting temperature in transition
metals, such as Ni, Fe, Co.

Computer simulations give us a new information about system, such
as coordinates of all particles of modelling system, which can not
be obtained by means of usual experimental methods. However, using
atomic coordinates to calculate the pair correlative function
(PCF) or structure factors to compare them with experimental data
is not an effective method for investigating the liquid structure.
One of the best prospective approach to solving such tasks is the
Voronoi polyhedra (VP) and Delaunay simlices (DS) method
\cite{Medvedev1, Medvedev2}.

Computer simulation methods allow us to advance in understanding
the atomic structure of the amorphous and liquid state of matter.
But the main problem of methods, such as molecular dynamic (MD) or
Monte-Carlo, is a choice of adequate potentials of the
interactions between atoms. Usually, the disparities of the
computer and experimental results are connected with a
non-correspondence of used potentials.

Generally, the task of finding the exact and universal interatomic
potentials is quite difficult. Therefore, inverse methods
\cite{Reatto, Schommers}, in which the effective pair potentials
are derived from the experimental structural data, are one of the
best methods for obtaining the effective pair potentials and
computing structure, corresponding with real matter. Moreover, the
deriving potentials are effective and can be considered as
potentials that include many-particle correlation, which is
typical for liquid state.

In this work we used the self-consistence inverse method
\cite{Schommers} for obtaining the effective pair potentials of
liquid nickel at different temperatures above its melting point
from the experimental structural data \cite{Waseda}. The
potentials allow us to get an adequate structure of liquid nickel,
carry out Voronoi-Delaunay statistical-geometrical analysis of
local atomic structure and investigate its temperature evolution.

\end{section}

\begin{section}{Formalism}

\begin{subsection}{Self-consistent method for deriving the effective pair potentials}

In this study the predictor-corrector method, originally proposed
by Schommers \cite{Schommers}, was used as an accurate method for
solving the inverse problem. This self-consistent method is based
on the integral equations theory and computer simulations. The
procedure of deriving $\varphi(r)$ is iterative. The
zero-approximation of the effective potential called predictor, is
given by:

\begin{equation}\displaystyle
\varphi^0(r)= -k_bT\ln{g^{exp}(r)},
\end{equation}
where $g^{exp}(r)$ is the experimental PCF, $k_b$ - is Boltcman
constant and $T$ is the absolute temperature of the system. Atomic
configuration, simulated by means of MD with known potential
$\varphi^0(r)$, corresponds to the pair correlation function
$g^{\tau-1}(r)$. The next iteration, called corrector, constructs
the corrected potential $\varphi^{\tau}(r)$:

\begin{equation}
 \varphi^\tau(r)=
 \varphi^0(r)-k_bT\ln{\frac{g^{exp}(r)}{g^{\tau-1}(r)}}.
\end{equation}

The agreement between experimental and model PCFs is improved with
each next iteration. The iterative process is repeated until the
discrepancy ~$\chi$ between the criterion (experimental) and
simulated PCFs becomes smaller than the desired accuracy:

\begin{equation}
\chi=\frac{1}{n_1-n_2}\sum[g^{exp}(r)-g^\tau(r)]^2,
\end{equation}
where $n_1$ and $n_2$ - are the numbers of initial and final
points of histogram, respectively.

\end{subsection}

\begin{subsection}{Voronoi-Delaunay statistical-geometrical analysis}

Models of dense disordered systems of spherical atoms have
definite statistical motives in the arrangement of these
structural elements. Unlike crystals, in which the structural
motives are described in terms of the translation order, in this
case the methods need to be different from those in
crystallography. Thus, Voronoi-Delaunay structural analysis was
carried out for describing the local atomic configurations as
function of the atomic coordinates.

The most widely applicable forms of the statistical-geometrical
analysis of random systems by VP are topological (distribution of
the number of faces and number of edges per face and topological
types) and metric (distribution of VP volumes, face areas, the
sphericity coefficient etc.) properties \cite{Medvedev1}. The
distribution of the coefficient of the sphericity $K_{sph}$ of VP
was chosen in the present work, where the sphericity coefficient
is defined by the expression:
\begin{equation}
K_{sph}=\frac{36\pi V^2}{S^2}.
\end{equation}
Here, $V$ is the volume, and $S$ is the surface area of given
polyhedron. Thus, this measure defines a deviation of the shape of
VP from regular sphere and characterizes homogeneity of atomic
environment.

It is well known, that packing of four spherical particles in
vertexes of regular tetrahedron is the most energetically
advantageous atomic configuration. Since each DS is a tetrahedron,
it is convenient to choose the deviation of the simplex shape from
an ideal tetrahedron as a quantitative measure of the DS. So, the
measure of tetrahedricity, ~$Tr$, we used was proposed in
\cite{Medvedev2}:

\begin{equation}\displaystyle
Tr=\frac {\displaystyle \sum(l_i-l_j)^2}{15<l>^2},
\end{equation}
where $l_i$ is the length of the $i$-th edge; $<l>$ is the average
length for this simplex. This measure was constructed to be zero
for an ideal tetrahedron and to increase with distortion. The
relative weight of tetrahedricity, $Tr_b=0.018$, is taken from
\cite{Medvedev2}, where it was proposed as boundary value for
perfect tetrahedra.

However the most interesting features of disordered systems are
seen in the space structural correlation. Since, there is unique
correspondence between the Voronoi and Deluanay networks, so using
$Tr$ as quantitative measures of DS each vertex of Voronoi network
can be assigned a value equal to this measure. To discriminate
aggregates of atoms with $Tr$ surrounding them it is convenient to
use coloring. In this case only vertices which have values less
than the chosen boundary value are colored. Thus, the analysis of
the spatial arrangement of atoms with different structural
patterns in the  structure of liquid reduces to the analysis of
typical graphs in  the Voronoi network.

\end{subsection}

\begin{subsection}{Time correlative Van Hove function}

Time correlation Van Hove function describes the temporal
relaxation of atoms local order and allows us to determine the
local order destruction time. The Van Hove function determines the
number of particles $N(r,\delta t)$ at the time $\delta t$, were
situated on the distance interval from $r$ to $r+dr$, under the
condition, that the particle was at the center point at the
initial time and $\Omega_0$ is the particle's volume:

\begin{equation}
\displaystyle Gd(r,\delta t)=\Omega_0\frac{N(r,\delta t)}{4\pi
r^{2}dr}.
\end{equation}

However, it would be interesting to examine the Van Hove function
only for particles that remain in the tetrahedral formations
during time interval $\delta t$. So, the function allows us to
estimate the life time of such atomic configurations.

\end{subsection}
\end{section}

\begin{section}{Simulation procedure}

\begin{subsection}{ Effective pair potentials}

In this work the experimental PCFs, $g^{exp}(r)$, obtained for
liquid nickel at $1773$, $1873$, $1923$, $2023 K$ \cite{Waseda},
are used as criterion functions in the inverse method. The derived
effective pair potentials are shown in fig. 1. They have some
variations for the four temperatures and display long-ranged
oscillations, that can be associated with Friedel's oscillations.
In our study performed in the microcanonical ensemble, we used
molecular dynamical models with $N=1688$ spherically symmetric
particles, arranged in a cubical box with the density from
$0.07921$ to $0.07942$ \AA$^{-3}$, corresponding to investigated
temperatures \cite{Smithells}. The MD computations, based on the
effective potentials $\varphi(r)$, reproduce the PCFs, which are
given in insertions on fig.1. The deviation $\chi$ between the
calculated and experimental functions reaches $10^{-4}$.

The good agreement between the computed and experimental
structural properties allows us to use Voronoi-Delaunay analysis
of liquid nickel's local structure at different temperatures.

\end{subsection}
\begin{subsection}{Local structure analysis}

Distributions of the sphericity coefficient for different
temperatures are presented on the fig. 2. Asymmetrical
$K_{sph}$-distribution, with a most probable value $K_{sph}=0.712$
and a shoulder at $K_{sph}=0.69$, corresponds to the temperature
$T=1773 K$. A contribution to the main maximum forms 12-14 facets
VP which have a prevalent content of 5-edged facets. Polyhedron in
shape of the dodecahedron (formed from twelve regular 5-edged
facets) corresponds to full icosahedral atom packing. So, atoms,
which correspond to VP with value $K_{sph}=0.712$ and have edge
numbers higher then 12, can only correspond to segments of dense
icosahedral packing. We can suggest, that two competitive types of
VP exist in the liquid nickel's structure at $T=1773 K$.

$K_{sph}$-distributions, corresponding to $T>1773 K$, became
symmetrical with maxima position at $K_{sph}=0.69$. This value
coincide with shoulder position for the $K_{sph}$-distribution at
$T=1773K$. At the same time, in these structures there are no VP
having a several bordering regular 5-edged facets. This fact
confirms that MD-models at $T>1773K$ do not contain sufficiently
\mbox{large segments of icosahedral atom packing.}


Investigation of the space structural correlations on the Voronoi
network by means of the vertex coloring under condition $Tr<Tr_b$,
has shown, that tetrahedral configurations tend to unit into
spatial formations ($Tr$-clusters) with a regular simplicial
surrounding. However, properties of these clusters, such as size,
local order, life time are changed with temperature rise.

Fig. 3a demonstrates typical tetrahedral formations in the
structure at $T=1773 K$. The size of $Tr$-clusters at this
temperature reaches $5-10$ \AA. On the Voronoi network such
clusters are mainly composed by 5-membered graphs bounded by the
mutual edge. This graph corresponds to the atomic configuration
with 5-folder symmetry in shape of the pentagonal biprism. So, in
the simulated liquid nickel's structure at $T=1773 K$ the
$Tr$-clusters are represented by various polytetrahedral atomic
configurations with 5-folder local symmetry with an icosahedral
configuration among them.
$K_{sph}$-distributions for the liquid nickel
configurations.

At temperatures $T>1773 K$ the structure of tetrahedral formations
is simplified. $Tr$-clusters become isolated and compact, their
size reaches as little as $5-6$ \AA. Typical element of these
formations is presented by 5-membered graphs.

\end{subsection}

\begin{subsection}{Van Hove function}

Advantageous energetic atomic disposition in the $Tr$-clusters
allows us to expect, that such formations should be time stable.
We checked this suggestion with help of analysis of time
correlative Van Hove functions. The Van Hove functions, calculated
at $T=1773 K$, are shown on fig. 4a. The first and the second
shells of the neighbors in the $Tr$-clusters are preserved at
least during $\bigtriangleup t=10^{-11}s$, that is larger than
thermal fluctuation's time.

Fig. 4 ($b$, $c$ and $d$) displays the Van Hove functions,
calculated at temperatures $T>1773 K$. Only the first coordination
shell is preserved during $\delta t=5 \cdot 10^{-12}s$. We
suggest, that it is connected on the one hand, with increasing of
the dynamic mobility of particles with temperature rise, in the
other hand, with an absence of pronounced icosahedral segments in
liquid structure at high temperatures.

\end{subsection}
\end{section}

\begin{section}{Conclusions}

In the work the effective pair potentials were derived from the
experimental data for liquid nickel at four temperatures. Good
agreement between computed and experimental structural properties
of the simulated system has made it possible to perform
Voronoi-Delaunay analysis of local structure.

The results of these simulations allow us to draw a conclusion
that in liquid nickel's local structure in the vicinity of the
melting point there are spatial polytetrahedral formations with
local 5-folder symmetry (with an icosahedron among them, as the
best-known candidate for a basic structural unit of liquid). The
results obtained complement a well-known point of view about the
existence of the icosahedral short-range order in liquid metals,
which does not depend on their pre-melting structural
type~\cite{Schenk}. With temperature rise tetrahedral formations
evolve to isolate and compact ones with preserving of the local
five-folder symmetry. At the same time their life time is
decreased due to the increase of diffusion atom motions.

The appearance of the spatial formations in liquid structure can
leads to the essential change of structural viscosity
\cite{Frenkel}. Existent results of the experimental
investigations of the transition metals alloys of iron group ($Fe,
Co$) \cite{Lad'yanov1, Lad'yanov2} demonstrate the uneven
decreasing of the alloys viscosity at rising temperature. This
doesn't contradict the above mentioned arguments. These facts can
be the evidence of the structural transitions, which exist in the
alloys with rising temperature. Hence, one can expect that the
nickel melt will exhibit the ordinary Newton-liquids properties,
since the structural clusters are very small. However, the
clusters size and their concentration growth with temperature
decrease. Because of these processes the melt can have
structured-liquid properties, in particular, viscosity dependence
on flow velocity gradient can appear.
\end{section}

Figures:

Fig.~1:The effective pair potentials of liquid nickel at different
temperatures $1773K$, $1823K$, $1923K$, $2023K$ and corresponding
pair correlation functions ( circles - experimental data, solid
lines - simulated data).

Fig.~2:$K_{sph}$-distributions for configurations of liquid nickel
at different temperatures.

Fig.~3:Tetrahedral formations on the Voronoi network (thick lines)
and corresponding atomic configurations (spheres) a) $T=1773K$:
The dodecahedron on the Voronoi network corresponds to full
icosahedral atomic configuration; b) $T=2023K$: $Tr$-clusters
became isolated and compact, with typical element as the
pentagonal biprism.

Fig.~4:Van Hove functions $Gd(r,\delta t)$, calculated for atoms,
which remain in the $Tr$-clusters during the time interval $\delta
t$.

%
%
%
%
%


\begin{thebibliography}{99}

\bibitem{Baum} B.A. Baum, G.A. Khasin, G.V. Taygunov, et al., Liquid Steel
(Mettalurgiya, Moskow, 1984);

\bibitem{Vatolin} N.A. Vatolin and E.A. Pastukhov, Diffraction studies of the
Structures of High-Temperature Melts (Nauka, Moscow, 1980);

\bibitem{Schenk} T. Schenk, D. Holland-Moritz, V. Simonet, R. Bellissent, D.M.
Herlach, Phys. Rev. Lett. 89, 075507 (2002);

\bibitem{Frank} F. C. Frank, Proc. R. Soc. London A 215, 43 (1952);

\bibitem{Medvedev1} N.N. Medvedev and Y.I. Naberukhin, J.Computer Phys. 67, 223
(1986);

\bibitem{Medvedev2} N.N. Medvedev, J.Phys. Condens.Matter, 2, 9145 (1990);

\bibitem{Reatto} L. Reatto , D. Levesque , J.J. Weis, Phys. Rev. A 33, 3451 (1986);

\bibitem{Schommers} W. Schommers, Phys. Rev. A 28, 3599 (1983);

\bibitem{Waseda} Y. Waseda, The Structure of Non-Crystalline Materials (McGraw-Hill, New York, 1980), an electronic database can be found at
http://www.iamp.tohoku.ac.jp/database/scm/index.html;

\bibitem{Smithells}C.J. Smithell, Smithells Metals reference book (Butterworths, London, 1983);


\bibitem{Frenkel} Ya.I. Frenkel, Kinetic Theory of Liquids (Nauka, Leningrad,
1975);

\bibitem{Lad'yanov1} V.I. Lad'yanov, A.L. Beltyukov, et al., JETP Let. 72, 6,
301 (2000);

\bibitem{Lad'yanov2} V.I. Lad'yanov and A.L. Beltyukov, JETP Let. 7,
2, 128 (2000);

\end{thebibliography}
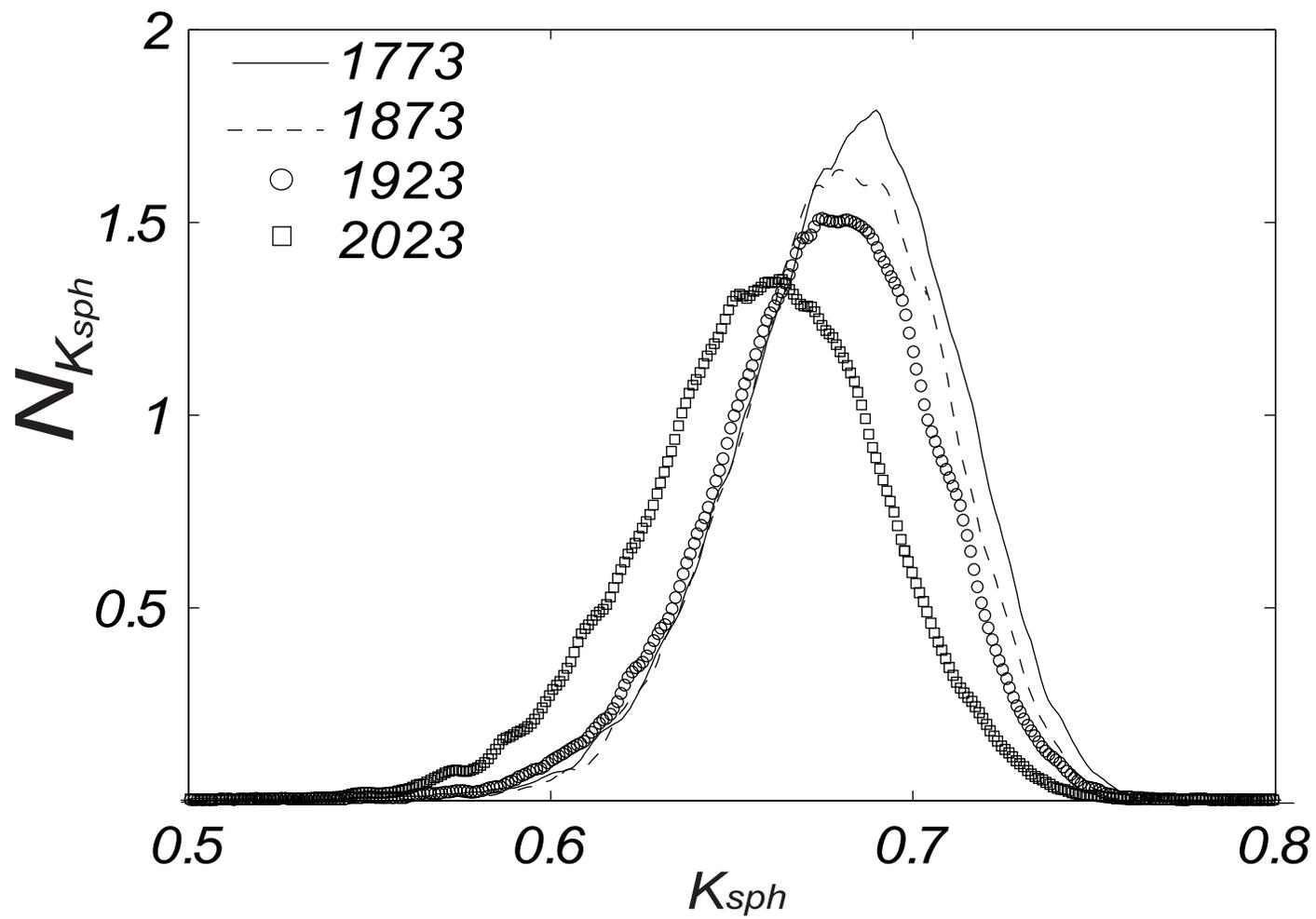
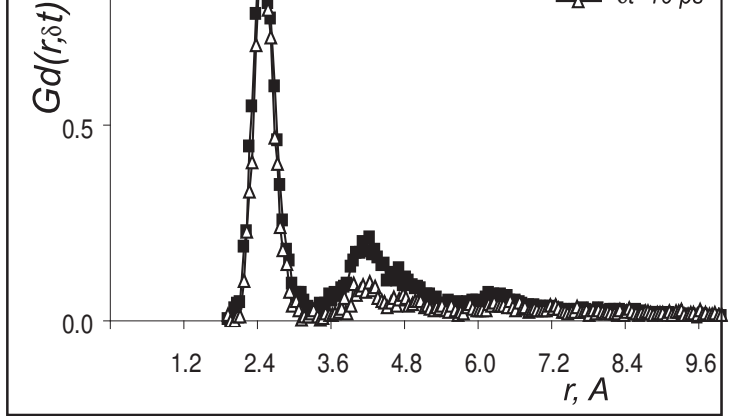
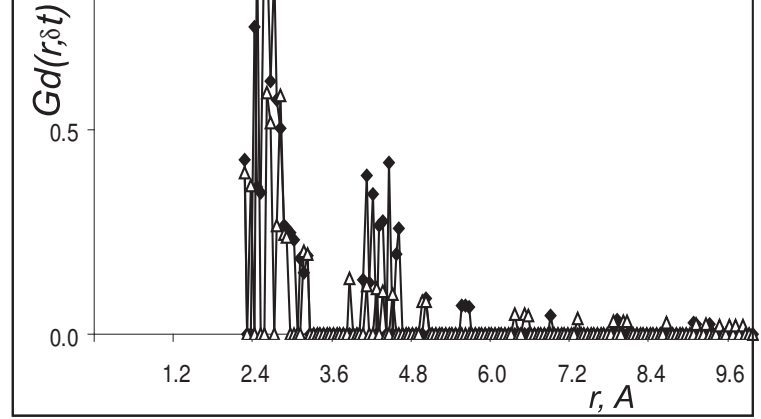
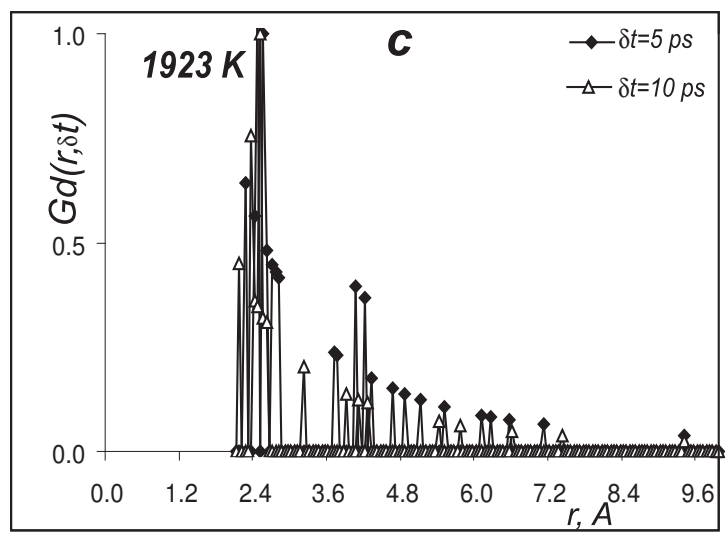
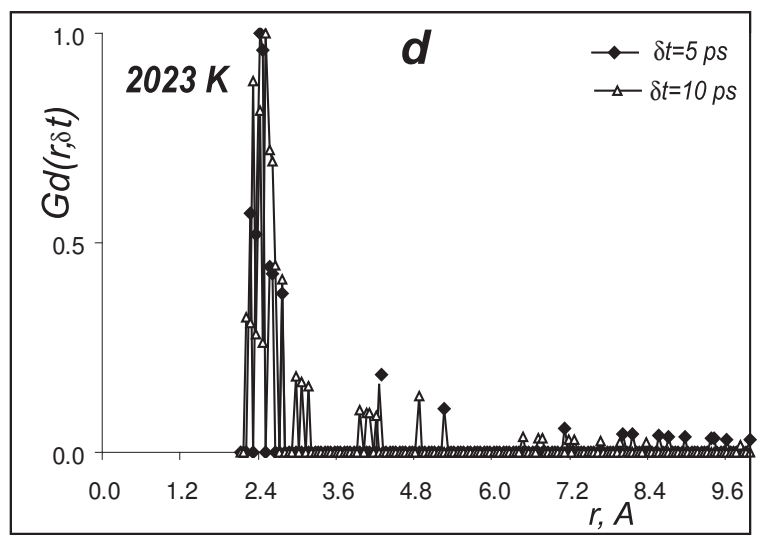
\end{document}